\documentclass[journal,comsoc,10pt]{IEEEtran}
\usepackage{acronym}
\usepackage{amsfonts}
\usepackage[dvips]{graphicx}
\usepackage{times}
\usepackage{cite}
\usepackage{amsmath}
\usepackage{array}
\usepackage{amssymb}
\usepackage{stfloats}
\usepackage{diagbox}
\usepackage{graphicx}
\usepackage{footnote}
\usepackage{amsthm}
\usepackage{booktabs}
\usepackage{array}
\usepackage{algorithmic}
\usepackage[ruled,vlined]{algorithm2e}
\usepackage[ruled]{algorithm2e}
\usepackage{subeqnarray}
\usepackage{cases}
\usepackage{threeparttable}
\usepackage{color}
\usepackage{epstopdf}
\usepackage{multirow}
\usepackage{tabularx}
\usepackage{enumerate}
\usepackage{multicol}
\usepackage{hyperref}
\usepackage{subfig}
\usepackage{caption}
\def\bb0{{\mathbb{0}}}

\def\bb{{\mathbf{b}}}

\def\bd{{\mathbf{d}}}

\def\bm{{\mathbf{m}}}
\def\bn{{\mathbf{n}}}

\def\br{{\mathbf{r}}}
\def\bs{{\mathbf{s}}}

\def\bx{{\mathbf{x}}}
\def\by{{\mathbf{y}}}

\def\b0{{\mathbf{0}}}

\def\bA{{\mathbf{A}}}

\def\bD{{\mathbf{D}}}
\def\bE{{\mathbf{E}}}

\def\bH{{\mathbf{H}}}
\def\bI{{\mathbf{I}}}

\def\bW{{\mathbf{W}}}

\def\bbC{{\mathbb{C}}}

\def\bbE{{\mathbb{E}}}

\def\sf0{{\mathsf{0}}}

\def\rmd{{\mathrm{d}}}

\def\rmu{{\mathrm{u}}}

\def\rm0{{\mathrm{0}}}

\def\Nr{{N_\mathrm{R}}}

\def\Nr{{N_{\mathrm{r}}}}

\acrodef{CSI}[CSI]{channel state information}
\acrodef{CSIT}[CSIT]{channel state information at the transmitter}
\acrodef{CSIR}[CSIR]{channel state information at the receiver}
\acrodef{MIMO}[MIMO]{multiple-input multiple-output}
\acrodef{SISO}[SISO]{single-input single-output}
\acrodef{MISO}[MISO]{multiple-input single-output}
\acrodef{SIMO}[SIMO]{single-input multiple-output}
\acrodef{ADCs}[ADCs]{analog-to-digital convertors}
\acrodef{SNR}[SNR]{signal-to-noise ratio}
\acrodef{AWGN}[AWGN]{additive white Gaussian noise}
\acrodef{MRT}[MRT]{maximal ratio transmission}
\acrodef{DFT}[DFT]{Discrete Fourier Transform}
\acrodef{ULA}[ULA]{uniform linear array}
\acrodef{UPA}[UPA]{uniform planar array}
\acrodef{LS}[LS]{least squares}
\acrodef{ALMMSE}[ALMMSE]{approximate linear minimum mean squared error}
\acrodef{QIHT}[QIHT]{quantized iterative hard thresholding}
\acrodef{QIST}[QIST]{quantized iterative soft thresholding}
\acrodef{SVD}[SVD]{singular value decomposition}

\usepackage{bm}

\begin{document}

\title{Model-Driven Deep Learning for  Massive MU-MIMO with Finite-Alphabet Precoding}
\author{Hengtao He,~\IEEEmembership{Student Member,~IEEE,}
Mengjiao Zhang,
Shi Jin,~\IEEEmembership{Senior Member,~IEEE,} \\
Chao-Kai Wen,~\IEEEmembership{Member,~IEEE,}
and Geoffrey Ye Li,~\IEEEmembership{Fellow,~IEEE}

\thanks{Manuscript received Feb 22, 2020; revised Apr 17, 2019 and May 13, 2020; accepted Jun 8, 2020. The work was supported in part by the National Key Research and Development Program 2018YFA0701602, the National Science Foundation of China (NSFC) for Distinguished Young Scholars with Grant 61625106, and the NSFC under Grant 61941104. The work of H. He was supported in part by the Scientific Research Foundation of Graduate School of Southeast University under Grant YBPY1939 and the Scholarship from the China Scholarship Council under Grant 201806090077. The work of C.-K. Wen was  supported  in  part  by  the  Ministry  of  Science  and  Technology  of  Taiwan  under  grants  MOST 108-2628-E-110-001-MY3. The associate editor coordinating the review of this paper and approving it for publication was Prof. D. Ciuonzo.
\emph{(Corresponding author: Shi Jin.)}}

\thanks{H.~He, M. Zhang, and S.~Jin are with the National Mobile Communications Research
Laboratory, Southeast University, Nanjing 210096, China (e-mail: hehengtao@seu.edu.cn, mjzhang@seu.edu.cn, and jinshi@seu.edu.cn).}
\thanks{C.-K. Wen is with the Institute of Communications Engineering, National
Sun Yat-sen University, Kaohsiung 804, Taiwan (e-mail: chaokai.wen@mail.nsysu.edu.tw).}
\thanks{G.~Y.~Li is with the School of Electrical and Computer Engineering,
Georgia Institute of Technology, Atlanta, GA 30332 USA (e-mail:
liye@ece.gatech.edu).}
}

\maketitle

\begin{abstract}
Massive multiuser multiple-input multiple-output (MU-MIMO) has been the mainstream technology in fifth-generation wireless systems. To reduce high hardware costs and power consumption in massive MU-MIMO, low-resolution digital-to-analog converters (DAC) for each antenna and radio frequency (RF) chain in downlink transmission is used, which brings challenges for precoding design. To circumvent these obstacles, we develop a model-driven deep learning (DL) network for massive MU-MIMO with finite-alphabet precoding in this article. The architecture of the network is specially designed by unfolding an iterative algorithm. Compared with the traditional state-of-the-art techniques, the proposed DL-based  precoder shows significant advantages in performance, complexity, and robustness to channel estimation error under Rayleigh fading channel.
\end{abstract}


\IEEEpeerreviewmaketitle
\vspace{-0.5cm}
\section{Introduction}
Massive multiuser multiple-input multiple-output (MU-MIMO) wireless systems, where the base station (BS) is equipped with several hundreds of antenna elements, have significant improvements in spectral efficiency, energy efficiency and reliability \cite{massivemimo}. Increasing the number of RF chains at the BS could, however, result in significant increases in hardware costs and power consumption. Therefore, practical massive MU-MIMO systems may require low-cost and power-efficient hardware components at the BS. Specifically, in downlink transmissions, equipping low-resolution digital-to-analog converters (DACs) can greatly reduce the cost and power consumption \cite{DACpaper} but directly limits the degree of freedom for the output signals and brings challenges into precoder design.

To address the issues, many efficient quantized precoding schemes have been proposed \cite{SQUID,C2PO,IDE,shao}. By using biconvex relaxation, an one-bit precoding algorithm for massive MU-MIMO systems is developed in \cite{SQUID}, which outperforms the zero-forcing (ZF) precoder directly followed by quantization, but still with high complexity. To reduce the complexity, a computationally-efficient one-bit beamforming algorithm referred to as C2PO, and its VLSI architectures are proposed in \cite{C2PO}.  Aside from one-bit DACs,  a universal algorithm for a downlink massive MU-MIMO system with finite-alphabet precoding in \cite{IDE} includes low-resolution DACs and phase-shifter-based architecture, which presents excellent performance and can be implemented with low computational complexity.

Recently, deep learning (DL) has been applied to physical layer communications \cite{Modeldriven18DL,DL2018Qin,DL2OFDM},
such as channel state information (CSI) feedback \cite{CSINet}, channel estimation \cite{DL2018HE} and precoder design \cite{BFNN,DLbeam}. However, most existing DL-based precoders focus on data-driven approaches, which consider the precoder as a black box and train it by using a huge volume of data. By contrary, model-driven DL approaches \cite{Modeldriven18DL} significantly reduce required volume of data for training and converge fast.

A neural-network optimized precoder in \cite{C2PO-Net}, named NNO-C2PO, recently has been proposed for one-bit precoding. The NNO-C2PO is obtained by unfolding the biConvex one-bit PrecOding (C2PO) algorithm and optimizing several manual parameters. Unfolding iterative algorithm into neural network has first been proposed for sparse signal recovery \cite{TISTA} and  become a promising  model-driven DL technique for wireless communications \cite{Modeldriven18DL, OAMP-Net2}. Inspired by these works, we unfold the iterative discrete estimation (IDE2) \cite{IDE} precoder into a neural network and add some adjustable parameters to learn the step size and damping factor to develop a model-driven DL approach for massive MU-MIMO with finite-alphabet precoding.

\emph{Notations}---For any matrix $\bA$, $\bA^{T}$, $\bA^{H}$, and ${ \mathrm{tr}}(\bA)$ denote the transpose,  conjugated transpose, and  trace of $\bA$, respectively. In addition, $\mathbf{I}$ is the identity matrix, $\mathbf{0}$ is the zero matrix. A proper complex Gaussian with mean $\boldsymbol{\mu}$ and covariance $\boldsymbol{\Theta}$ can be described by the probability density function,
\begin{equation*}
  \mathcal{N}_{\mathbb{C}}(\mathbf{z};\boldsymbol{\mu},\boldsymbol{\Theta})=\frac{1}{\mathrm{det}(\pi \boldsymbol{\Theta})}
  e^{-(\mathbf{z}-\boldsymbol{\mu})^{H}\boldsymbol{\Theta}^{-1}(\mathbf{z}-\boldsymbol{\mu})}.
\end{equation*}

\begin{figure*}
  \centering
  \includegraphics[width=12cm]{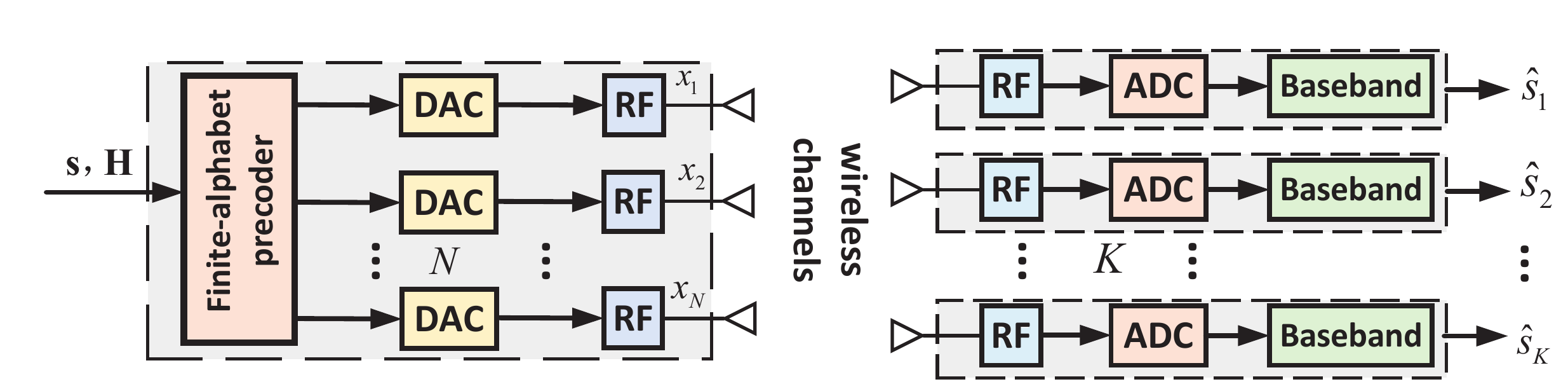}
  \caption{.~~The diagram of massive MU-MIMO downlink system with low-resolution DACs. The BS uses a finite-alphabet precoder and $N$ antennas to serve $K$ users.}\label{JCD}
\end{figure*}\label{DACfig}
\section{System Model and Problem Formulation}\label{JCD_section}
In this section, we will first present the massive MU-MIMO downlink system model. Then, the finite-alphabet precoding is formulated as an integer programming problem.

\subsection{System Model}

Consider downlink transmission of a narrow-band massive MU-MIMO system illustrated as Fig.\,1, where a BS equipped with $N$ antennas serves $K$ single-antenna users. In the figure,  $\mathbf{s}\in\bbC^{K}$ is the required transmitted constellation points for $K$ users. We assume $\bbE\{\bs\bs^{H}\} = \mathbf{I}$ for the transmitted symbol vector. The received signal of all users $\by = [y_{1},y_{2},\ldots,y_{K}]\in \bbC^{K}$ can be obtained by
\begin{equation}\label{eqmodl}
  \by = \bH\bx+\bn,
\end{equation}
where $\bx=[x_{1},x_{2},\ldots,x_{N}]\in \bbC^{N}$ is the precoded signal transmitted from the BS and carries transmitted information for different users.  $\bn \sim \mathcal{N}_{\mathbb{C}}(0,\sigma^{2}\mathbf{I}_\Nr) $  is the additive white Gaussian noise (AWGN) vector and $\bH\in\bbC^{N\times K}$ is  the downlink MIMO channel matrix. We assume the entries of channel $\bH$ are independent circularly-symmetric complex Gaussian random variables with unit variance, i.e., $h_{i,j} \sim \mathcal{N}_{\bbC}(0,1)$. In downlink transmission, the precoded vector $\bx$ should satisfy the average power constraint
\begin{equation}\label{eqcon}
  \frac{1}{N}\bbE\{\|\bx\|_{2}^{2}\} \leq P_{t},
\end{equation}
where $P_{t}$ is the transmit power of each antenna.

Each user is assumed to be able to rescale the received signal by a precoding factor $\beta$ to obtain  the estimated constellation points $\hat{s}_{k}$ for $ k= 1,\ldots, K$, i.e., $\hat{s}_{k} = \beta y_{k}$, that is, $\hat{\mathbf{s}} = [\hat{s}_{1},\hat{s}_{2},\ldots,\hat{s}_{k}]^{T} = \beta\mathbf{y} = \beta\bH\bx+\beta\bn $. The precoding factor $\beta$ is determined by the precoder. We consider that the zero-forcing (ZF) precoder is used, the precoding factor $\beta$ is accordingly obtained by power constraint in (\ref{eqcon}) and  given by $\beta = \sqrt{\frac{\mathrm{tr}((\bH\bH^{H})^{-1})}{N P_{t}}} $ \cite{SQUID,IDE}. As a result,  the MSE of the estimation error between the rescaled received signal $\beta\by$ and the transmitted symbol vector $\bs$ can be written as
\begin{equation}\label{eqmse1}
  \bbE_{\bs} \left\{\|\bs - \beta \by \|^{2}_{2}\right\} = \bbE_{\bs}\left\{\|\bs - \beta \bH\bx \|^{2}_{2}\right\}+\beta^{2}K\sigma^{2},
  \end{equation}
where the metric for interuser interference (IUI) is defined by $ \mathrm{IUI} = \bbE_{\bs} \left\{\|\bs - \beta \bH\bx \|^{2}_{2}\right\}$.
The MSE in ($\ref{eqmse1}$) measures the performance of the precoder and can be used as the loss function for DL network.
\vspace{-0.5cm}
\subsection{Problem Formulation}
If the ZF precoder is used at the BS, the IUI can achieve zero if the BS is with infinite-resolution DACs and high-linearity power amplifiers \cite{IDE}. In practical setting, each antenna is equipped with a low-cost constrained RF chain, including low-resolution DACs, low-resolution analog phase shift to reduce the cost and power consumption.

In this article, we assume each entry of the vector $\bx$ is restricted to a finite-alphabet set $\bm{\chi} = \{\chi_{0},\ldots,\chi_{M-1}\}$, where $M = |\bm{\chi}|$. Such finite-alphabet set, including QAM and PSK symbols, can relax the perfect hardware requirement \cite{SQUID,DACpaper,C2PO,IDE}. However, it brings several challenges in precoder design, especially obtaining zero IUI becomes difficult in general. The goal of finite-alphabet precoding is to design a precoder $\mathcal{P}(\bs,\bH)$ that minimizes the MSE between the rescaled received signal $\beta\by$ and the transmitted symbol vector $\bs$ under the power constraint ($\ref{eqcon}$), which is  given by
\begin{align}\label{eqopt}
  \min \limits_{\bx,\beta} & \quad \bbE_{\bs} \{\|\bs - \beta \bH\bx \|^{2}_{2}\} \nonumber \\
  \mathrm{s.t.} & \quad \bx \in \bm{\chi}^{N}, \beta > 0
\end{align}

The problem of ($\ref{eqopt}$) is related to an integer programming problem, which is always NP-hard. Several iterative approaches have been proposed for finite-alphabet precoding, including SQUID \cite{SQUID}, C2PO\cite{C2PO}, IDE2\cite{IDE}, and NNO-C2PO\cite{C2PO-Net}. In the next section, we will introduce the IDE2 precoder as a representative algorithm and improve it by DL.
\section{IDE2-Net}\label{network}
After introducing the IDE2 precoding in \cite{IDE}, we propose IDE2-Net by unfolding the IDE2 precoder for finite-alphabet precoding. Then, we present the network architecture and elaborate the learnable variables and computational complexity of the IDE2-Net.

\subsection{IDE2 Precoder}\label{IDE2}
\begin{algorithm}[t]
 \textbf{Inputs}: $\bs$, $\tilde{\bH} = \beta\bH$\\
 \textbf{Initial}: $t=0$, $\bx_{\mathrm{d}}^1 = \textbf{0}$, $\alpha = 0.95$\\

 \While{$t < T$}
 {
 \begin{equation}\label{eq1}
  \bW_{\mathrm{u}} = [\mathrm{diag}(\tilde{\bH}^{H}\tilde{\bH})]^{-1} \tilde{\bH}^{H}
 \end{equation}
 \begin{equation}\label{eq2}
 \br^{t} =  \bx_{\mathrm{d}}^t + \gamma^{t} \bW_{\mathrm{u}} (\bs - \tilde{\bH}\bx_{\mathrm{d}}^t)
\end{equation}
 \begin{equation}\label{eq3}
 \bx^{t+1} = \Pi_{\chi}{\left( \br^{t} \right)}
 \end{equation}
 \begin{equation}\label{eq4}
 \bx_{\mathrm{d}}^{t+1} = \alpha \bx_{\mathrm{d}}^{t} +(1-\alpha)\bx^{t+1}
 \end{equation}

 $t \leftarrow t+1$}
 \textbf{Output} $ \bx =\bx^{t+1}$
 \caption{IDE2 algorithm}
\end{algorithm}

\begin{figure*}[t]
  \centering
  \includegraphics[width=12cm]{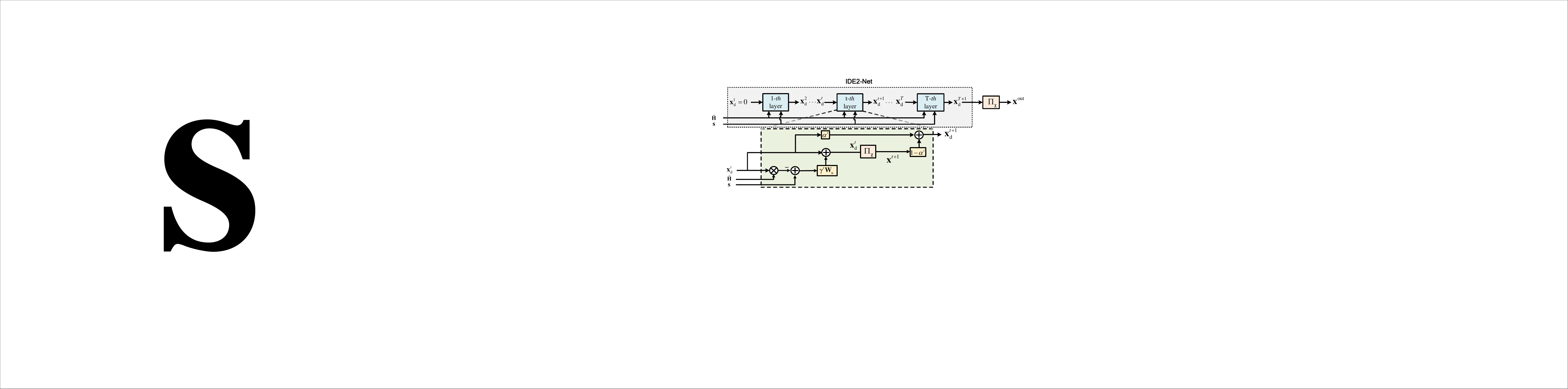}
  \caption{.~~The diagram of IDE2-Net precoder. The network consists of $T$ cascade layers and each layer has the same structure and contains the linear estimator $\mathbf{W}_{\rmu}$, nonlinear estimator $\Pi_{\chi}$, trainable variables $\gamma^{t}$ and  $\alpha^{t}$.}\label{IDE2-Net-fig}
\end{figure*}

The IDE2 preocder has been proposed to achieve finite-alphabet precoding for massive MIMO systems. The iterative procedure is illustrated in Algorithm $1$. The IDE2 precoder is with low-complexity and is mainly composed of two parts, linear estimator ($\ref{eq2}$) and nonlinear estimator ($\ref{eq3}$). Specifically, given the prior information of $\bx_{\mathrm{d}}^{t}$ on
\begin{equation}\label{eqprior}
  \bbE\{\bx\} = \bx_{\mathrm{d}}^{t} \quad \mathrm{and} \quad \bbE\{ (\bx-\bx_{\mathrm{d}}^{t})(\bx-\bx_{\mathrm{d}}^{t})^{H}\} = \frac{1}{\mu}\bI,
\end{equation}
the linear estimator in ($\ref{eq2}$) is an accurate approximation in massive MIMO systems for the optimal linear minimum mean-squared error (LMMSE) estimator
\begin{equation}\label{eqLMMSE}
  \br^{t} = \bx_{\mathrm{d}}^t + \bW\left(\bs - \tilde{\bH}\bx_{\mathrm{d}}^{t}\right),
\end{equation}
where
\begin{equation}\label{eqLMMSE2}
\bW = \left( \tilde{\bH}^{H}\tilde{\bH} + \mu\bI \right)^{-1}\tilde{\bH}^{H},
\end{equation}
and $\tilde{\bH} = \beta\bH$. From the perspective of estimation theory, the LMMSE estimator in ($\ref{eqLMMSE}$) is biased in each iteration. It can be revised  as an unbiased version by replacing $\bW$ with $\bW_{\rmu} = \bD\bW$, where $\bD$ is a diagonal matrix and the diagonal elements of
$\bW_{\mathrm{u}}\tilde{\bH} =\bD (\tilde{\bH}^{H}\tilde{\bH}+\mu\bI)^{-1}\tilde{\bH}^{H}\tilde{\bH} $ are 1. Therefore, $\bD$ is given by
\begin{equation}\label{eqdiag}
  \bD = [\mathrm{diag}(\bW\tilde{\bH})]^{-1}.
\end{equation}

In massive MIMO systems, the complexity for computing the LMMSE matrix ($\ref{eqLMMSE2}$) is very high. Fortunately, it can be addressed by using an approximation for the matrix inversion, which is given by
\begin{equation}\label{eqappro}
  \left( \tilde{\bH}^{H}\tilde{\bH} + \mu\bI \right)^{-1} \approx \frac{1}{\mu}\bI.
\end{equation}
Consequently, we obtain
\begin{equation}\label{eqwu}
  \bW_{\mathrm{u}} \approx [\mathrm{diag}(\tilde{\bH}^{H}\tilde{\bH})]^{-1}\tilde{\bH}^{H}.
\end{equation}
Thus, the linear estimator can be derived as ($\ref{eq2}$). As the BS is equipped with low-resolution DACs, the output of the linear estimator $\br^{t}$ should be projected onto finite-precoding set $\bm{\chi}$. The nonlinear estimator
can be denoted by

\begin{equation}\label{eqnonlinear}
  \bx^{t+1} = \Pi_{\chi}{\left( \br^{t} \right)}.
 \end{equation}

In each iteration, $\Pi_{\chi}$ generates $\bx^{t+1}$ exactly from the discrete point set $\bm{\chi}$, which is clearly different from $\br^{t}$. Therefore, a damping factor $\alpha$ should be introduced to help the algorithm to be convergent as follows,
\begin{equation}\label{eqdamping}
  \bx_{\mathrm{d}}^{t+1} = \alpha \bx_{\rmd}^{t} + (1-\alpha)\bx^{t+1}.
\end{equation}
The three equations in ($\ref{eq2}$)-($\ref{eq4}$) are executed iteratively until convergence.

\subsection{Network Architecture}
We propose a model-driven-DL-based precoder, named IDE2-Net, for massive MU-MIMO with finite-alphabet precoding. As Fig.\,\ref{IDE2-Net-fig} illustrated, the structure of the IDE2-Net is obtained by unfolding the IDE2 precoder and adding
several trainable parameters. The input of the IDE2-Net is the  signal vector $\bs$, and the channel matrix, $\tilde{\bH}$, and the final output is $ \bx^{\mathrm{out}} =\Pi_{\chi}{\left( \bx_{\mathrm{d}}^{T+1}  \right)}$. The IDE2-Net consists of $T$ cascade layers, where each has the same architecture but is with different trainable parameters.
For the $t$-th layer of the IDE2-Net, the calculative process is performed as follows
\begin{equation}\label{eq5}
  \bW_{\mathrm{u}} = [\mathrm{diag}(\tilde{\bH}^{H}\tilde{\bH})]^{-1} \tilde{\bH}^{H},
\end{equation}
\begin{equation}\label{eq6}
  \br^{t} =  \bx_{\mathrm{d}}^t + \gamma^{t} \bW_{\mathrm{u}} (\bs - \tilde{\bH}\bx_{\mathrm{d}}^t),
\end{equation}
\begin{equation}\label{eq7}
  \bx^{t+1} = \Pi_{\chi}{\left( \br^{t} \right)},
\end{equation}
\begin{equation}\label{eq8}
 \bx_{\mathrm{d}}^{t+1} = \alpha^{t} \bx_{\mathrm{d}}^{t} +(1-\alpha^{t})\bx^{t+1},
\end{equation}
where  $\bm{\theta}^{t} = \{{\gamma^{t},\alpha^{t}}\}$  are learnable variables in each layer.  When $ \gamma^{t} = 1 $ and $\alpha^{t} = 0.95$ for each layer, the IDE2-Net is reduced to the IDE2 precoder.  
%

In Section \ref{IDE2}, we have  introduced the principle of the IDE2 precoder. In fact, IDE2 appears nearly similar to the classical  projected gradient descent algorithm with the  objective function as illustrated in (\ref{eqopt}). The update for each iteration can be described as the following form
\begin{equation}\label{eqSGD}
  \bx_{\rmd}^{t+1} = \Pi_{\chi}{\left( \bx_{\mathrm{d}}^t + \lambda  \tilde{\bH} (\bs - \tilde{\bH}\bx_{\mathrm{d}}^t) \right)},
\end{equation}
where the step size $\lambda$ is always set to be sufficiently small to ensure convergence. Different from the classical projected gradient algorithm with a constant step size $\lambda$, the IDE2 precoder uses a vector form step size, $[\mathrm{diag}(\tilde{\bH}^{H}\tilde{\bH})]^{-1}$, that makes the linear estimator in ($\ref{eq2}$) unbiased. By contrary, the step size in (\ref{eqSGD}) is  $\lambda$, which makes each update in a biased manner. The step size in the IDE2 precoder is obtained by an approximation in a large system and is not the optimal one. Therefore, we will optimize the step size by tuning the parameter $\gamma^{t}$ with DL. Furthermore, we have found that the damping factor $\alpha$ is important in IDE2 precoder in \cite{IDE} and  $\alpha = 0.95$ is selected based on a trade-off between stability and speed of convergence. The optimized $\alpha$ depends on SNR and modulated symbols. Therefore, optimizing $\alpha$ through DL is an efficient approach.
\vspace{-0.2cm}
\subsection{Complexity Analysis}
Here, we compare the computational complexity of the proposed
IDE2-Net and other prior state-of-the-art methods, including SQUID \cite{SQUID}, IDE2 \cite{IDE}, C2PO \cite{C2PO}, NNO-C2PO \cite{C2PO-Net} in terms of the number of multiplication operations and learnable variables in each iteration, and summarize the corresponding results in Table \ref{table1}. The complexity of the IDE2, IDE-Net2 and SQUID is $O(2NK+N)$, which is much lower than that of C2PO and NNO-C2PO. Furthermore, the number of trainable variables of the IDE2-Net is only determined by the number of layers $T$ and is independent of the number of antennas $N$ and users $K$. This is a very attractive feature for massive MIMO systems. The few trainable variables can improve the stability and  convergence speed of the IDE2-Net in the training process.

\begin{table}[t]
\caption{Complexity analysis of different precoders}\label{table1}
  \centering
  \begin{tabular}{|c|c|c|}
  \hline
  Algorithm & Multiplications & Learnable Variables \\ \hline
  IDE2 & $O(2NK+N)$ & 0 \\ \hline
  IDE2-Net & $O(2NK+N)$ & 2 \\ \hline
  C2PO & $O(N^{2}K + NK)$ & 0 \\ \hline
  NNO-C2PO & $O(N^{2}K + NK)$ & 2 \\ \hline
  SQUID & $O(2NK+N)$ & 0 \\ \hline
\end{tabular}
\end{table}
\vspace{-0.2cm}
\section{Numerical Results}\label{Simulation}
In this section, we provide numerical results to show the performance of the proposed model-driven-DL-based precoder. First, we elaborate the implementation details and parameter settings. Then, the bit error-rate (BER) performance under i.i.d. Rayleigh MIMO channels\footnote{In this paper, we consider conventional Rayleigh MIMO channel, which is often used in \cite{SQUID, C2PO, IDE, shao}. The practical impairments considered in \cite{MLSP} can be investigated in our future work.} with perfect CSI is provided
. The robustness of IDE2-Net to channel estimation error is also investigated.
\vspace{-0.5cm}
\subsection{Implementation Details}
IDE2-Net is implemented by utilizing Tensorflow in our simulation by using a PC with GPU NVIDIA GeForce GTX 1080 Ti. The SNR is defined as
\begin{equation}\label{eqsnr}
  \mathrm{SNR}=\frac{NP_{t}}{\sigma^{2}},
\end{equation}
including the array gain and can be regarded as the received SNR for each user. We consider the extreme case that the BS is equipped with one-bit DACs. As existing deep learning APIs  are mostly devoted to process the real-valued data,  we consider equivalent real-valued representation for the system model (\ref{eqmodl}). The code will be available at
https://github.com/hehengtao/IDE2-Net.

The training data consists of a number of randomly generated pairs $\bd^{(i)}\triangleq(\bs^{(i)},\bH^{(i)})$. For each pair $\bd^{(i)}$, the channel $\bH$ is randomly generated from the i.i.d. MIMO channel model. The data $\mathbf{s}^{(i)}$ is generated from the $16$-QAM modulation symbol with $M$ being the  modulation order. We train the network with $1,000$ epochs. At each epoch, the training set contains $5,000$  different samples $\mathbf{d}^{(i)}$, and $1,000$  different validation samples. 
The network is trained using the stochastic gradient descent (SGD) optimizer. 
The learning rate is set
to be 0.1 initially, and has 10-fold decrease for every $200$ epochs, afterwards keeps $0.0001$ for the remaining epochs. The batch size is set to $100$.
We use the cost function defined as
\begin{equation}\label{eqmse}
  \mathrm{Loss}(\bm{\theta}) = \bbE_{\bs}\left\{\|\bs - \beta \bH\bx^{\mathrm{out}} \|^{2}_{2}\}\right.
  \end{equation}
where $\mathbf{x}^{\mathrm{out}}$ is the final output of the IDE2-Net. The nonlinear estimator $\Pi_{\chi}$ in each layer is not differentiable. We adopt the identity function in the backward pass, known as the straight-through estimator in \cite{stop_gradient} to decouple the forward-propagation and backpropagation. Then, all remaining operations in the IDE2-Net have well-defined gradients and can be differentiated automatically.
\begin{figure}[t]
  \centering
  \includegraphics[width=2.5in]{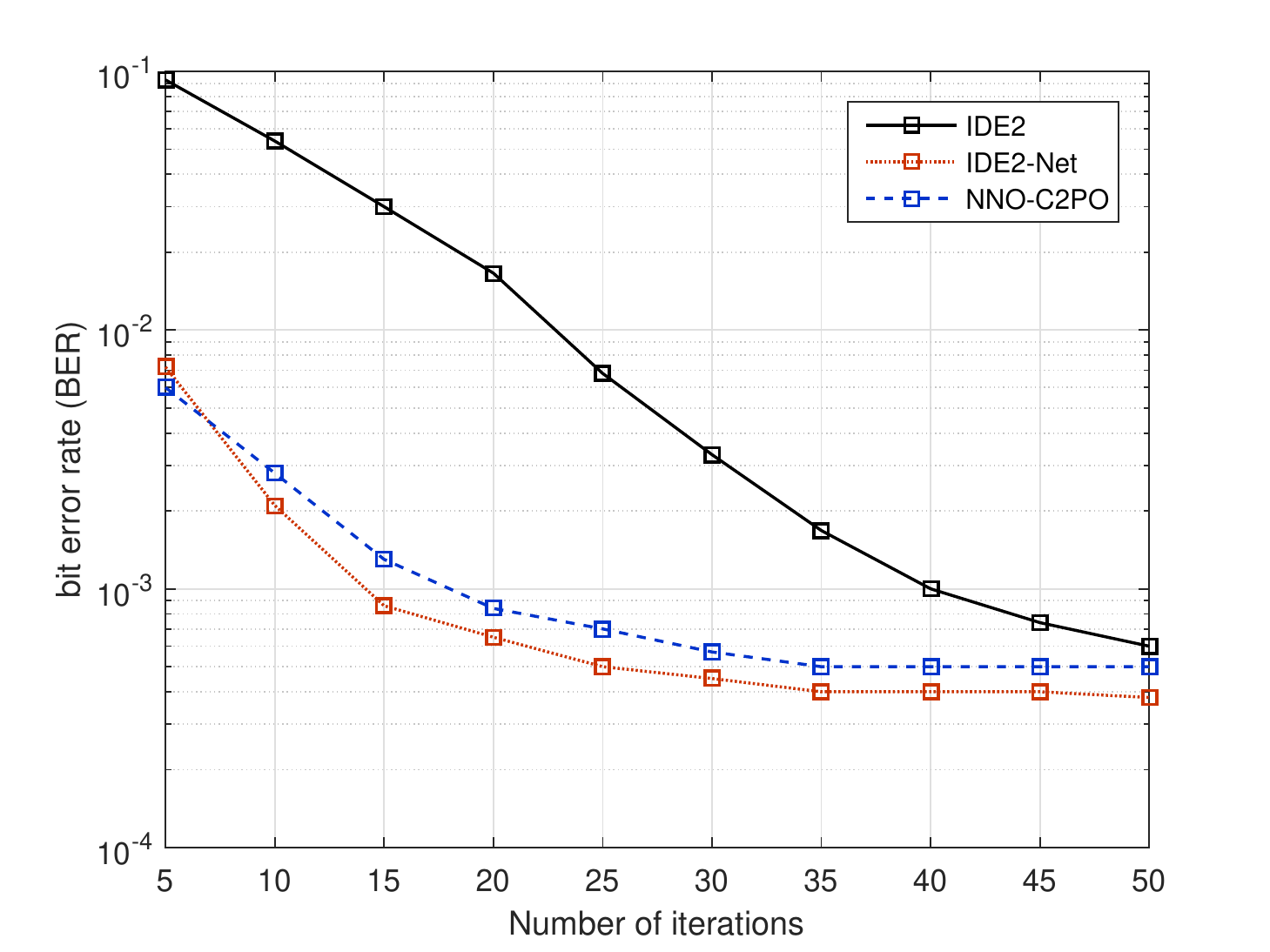}
  \caption{.~~Convergence analysis of IDE2-Net, IDE2, and NNO-C2PO precoder for a massive MU-MIMO system with $N = 128$ and $K=16$ in a Rayleigh-fading channel with SNR = $14$ dB.}\label{Fig41}
\end{figure}
\vspace{-0.5cm}
\subsection{BER Performance}
First, we analyze the convergence of the IDE2-Net, IDE2 and NNO-C2PO precoders. Fig.\,\ref{Fig41} illustrates the BER performance versus the number of iterations (layers). From the figure, the IDE2-Net has faster convergence speed than IDE2 and NNO-C2PO algorithms, which demonstrates the learnable variables can significantly accelerate convergence of the IDE2 algorithm. The reason is the IDE2-Net can learn the optimal parameters $\bm{\theta} = \{{\gamma^{t},\alpha^{t}}\}_{t=1}^{T}$ involved in the step size and damping factors from the data.

\begin{figure}[t]
  \centering
  \includegraphics[width=2.5in]{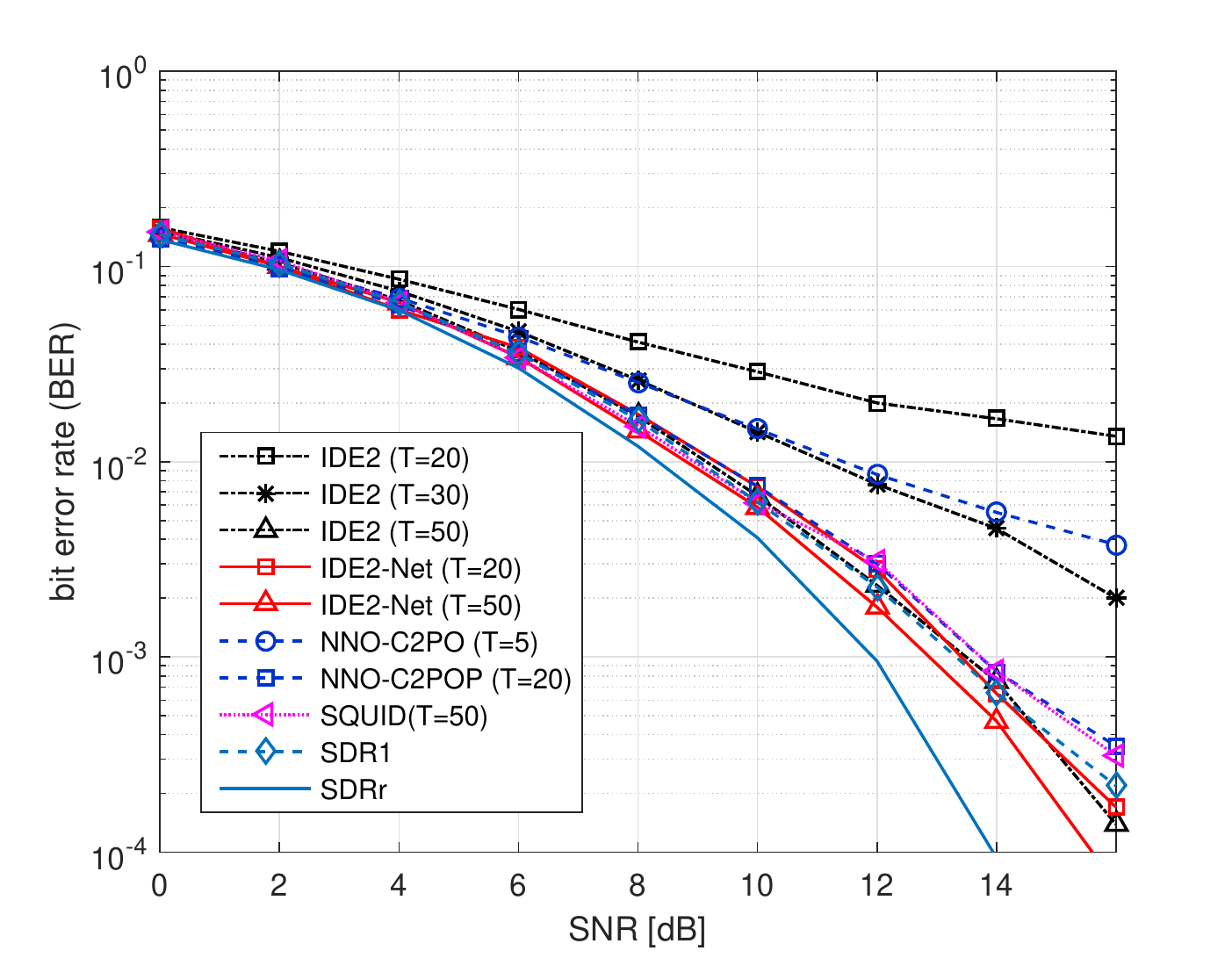}
  \caption{.~~BER performance of several precoders with different number of iterations for a massive MU-MIMO system with $N = 128$ and $K=16$ in a Rayleigh-fading channel.}\label{Fig1}
\end{figure}

Fig.\,\ref{Fig1} compares the BER performance of the IDE2-Net, IDE2\cite{IDE}, NNO-C2PO\cite{C2PO-Net}, SQUID, SDRr,  and SDR1 precoders in \cite{SQUID} with different numbers of iterations $T$. The NNO-C2PO is the state-of-the-art DL-based precoder. Except for SDRr, the IDE2-Net outperforms other precoders in terms of BER performance. However, the high complexity of SDRr precoder prevents its use for massive MU-MIMO systems with hundreds of antennas. Furthermore, compared with the IDE2 precoder, the IDE2-Net has significantly performance improvement for different numbers of iterations. However, the performance gain is decreased with the increase of the layers. This is because the IDE2-Net and IDE2 precoder will have similar performance when the number of layers is adequate. 
\subsection{Robustness to Channel Estimation Error}
In aforementioned sections, we have assumed that the BS has perfect CSI. Then, we investigate the robustness of the IDE2-Net to channel estimation error in this section. We assume the noisy channel is given by
\begin{equation}\label{eqHhat}
  \hat{\bH} = \sqrt{1-\epsilon}\bH + \sqrt{\epsilon}\bE,
\end{equation}
where $\epsilon \in [0,1]$ and $\bE$ has $\mathcal{N}_{\bbC}(0,1)$ entries. The value of $\epsilon = 0$, $\epsilon\in (0,1)$, and $\epsilon = 1$ correspond to cases with perfect CSI, partial CSI, or no CSI, respectively. We consider the IDE2-Net and NNO-C2PO are with $T=20$ layers and SNR = $14$ dB. The IDE2-Net and NNO-C2PO are trained with perfect CSI and deployed with imperfect CSI.

\begin{figure}[h]
  \centering
  \includegraphics[width=2.5in]{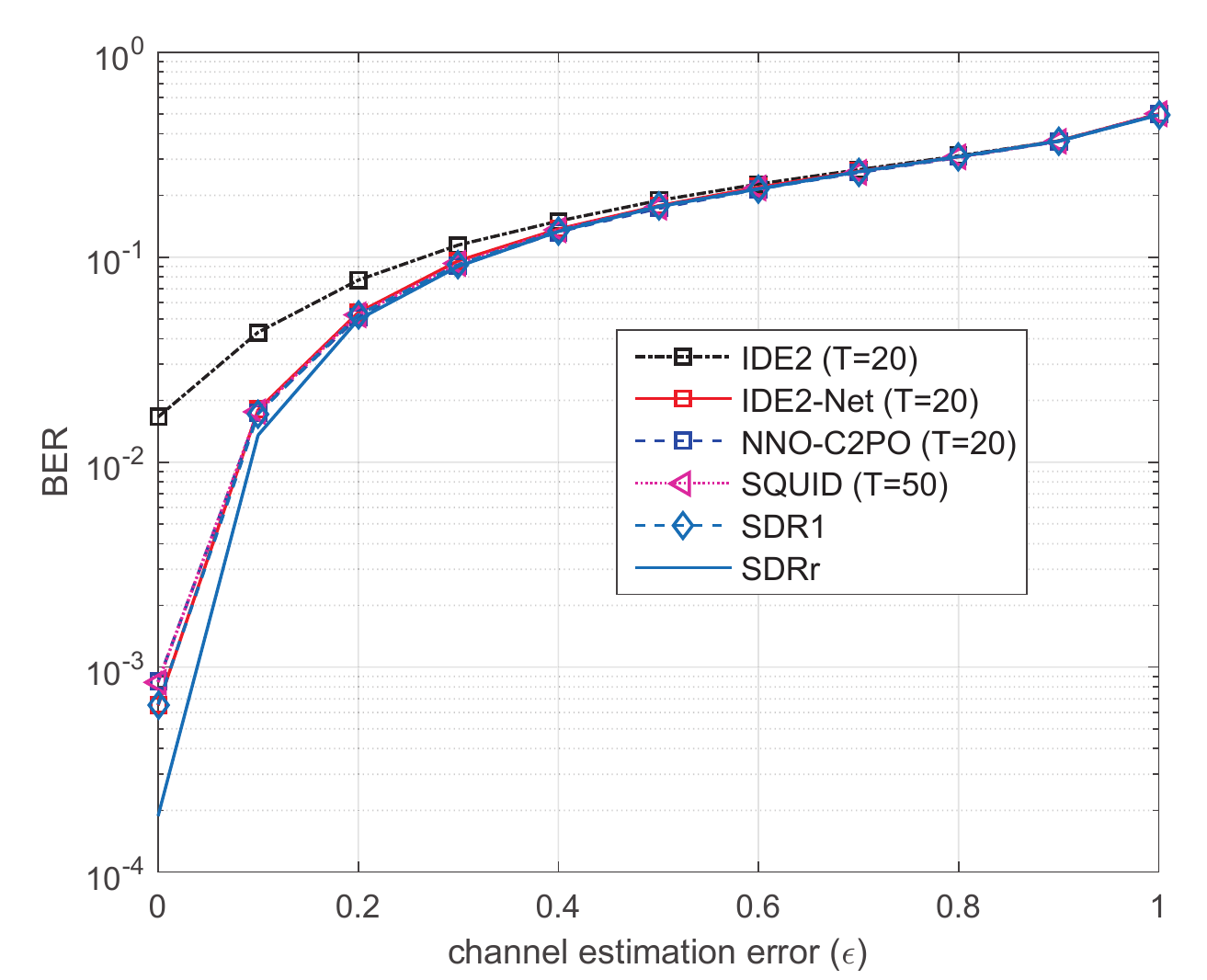}
  \caption{.~~BER performance of several precoders with channel estimation error $\epsilon$  with $N = 128$ and $K=16$ in a Rayleigh-fading channel with SNR =$14$ dB.}\label{Fig21}
\end{figure}

Fig.\,\ref{Fig21} demonstrates the robustness of different precoders to channel estimation error. As the figure illustrated, the robustness of IDE2-Net to channel estimation error is better than IDE2 precoder, and similar to NNO-C2PO, SQUID, and SDR1, which demonstrates the learnable variables can improve the robustness.
\vspace{-0.5cm}
\section{Conclusion}\label{con}
We have developed a model-driven DL network for massive MU-MIMO with finite-alphabet precoding, named IDE2-Net.  The IDE2-Net inherits the superiority of the iterative precoder and DL technique and presents excellent performance. The network has lower complexity than other precoding algorithms and only few adjustable parameters are required to be optimized. Simulation results demonstrate that significant performance gain can be obtained by learning corresponding optimal parameters from the data to improve the BER performance, accelerating convergence and enhancing robustness to channel estimation error with Rayleigh fading channel.


\begin{thebibliography}{10}

\bibitem{massivemimo}
T. L. Marzetta, ``Noncooperative cellular wireless with unlimited numbers
of base station antennas,'' {\em IEEE Trans. Wireless Commun.}, vol. 9,
no. 11, pp. 3590-3600, Nov. 2010.

\bibitem{DACpaper}
A. Mezghani, R. Ghiat, and J. A. Nossek, ``Transmit processing with low
resolution D/A-converters,'' in {\em Proc. IEEE Int. Conf. Electron., Circuits,
Syst. (ICECS)}, Yasmine Hammamet, Tunisia, Dec. 2009, pp. 683-686.

\bibitem{SQUID}
S. Jacobsson, G. Durisi, M. Coldrey, T. Goldstein, and C. Studer, ``Quantized precoding for massive MU-MIMO,''
{\em IEEE Trans. Commun.}, vol. 65, no. 11, pp. 4670-4684, Nov. 2017.

\bibitem{C2PO}
O. Casta\~{n}eda, S. Jacobsson, G. Durisi, M. Coldrey, T. Goldstein, and
C. Studer, ``1-bit massive MU-MIMO precoding in VLSI,'' {\em IEEE J.
Emerg. Sel. Topics Circuits Syst.}, vol. 7, no. 4, pp. 508-522, Dec. 2017.

\bibitem{IDE}
C.-J. Wang, C.-K. Wen, S. Jin, and S.-H. Tsai, ``Finite-alphabet precoding
for massive MU-MIMO with low-resolution DACs,'' {\em IEEE Trans. Wireless
Commun.}, vol. 17, no. 7, pp. 4706-4720, Jul. 2018.

\bibitem{shao}
M. Shao, Q. Li, W.-K. Ma, and A. M.-C. So, ``A framework for
one-bit and constant-envelope precoding over multiuser massive MISO
channels,'' {\em IEEE Trans. Signal. Process.}, vol. 67, no. 20, pp. 5309-5324,
Oct. 2019.

\bibitem{Modeldriven18DL}
H. He, S. Jin, C.-K. Wen, F. Gao, G. Y. Li, and Z. Xu, ``Model-driven deep learning for physical layer communications,'' \emph{IEEE Wireless Commun.},
vol. 26, no. 5, pp. 77-83, Oct. 2019.

\bibitem{DL2018Qin}
Z.-J. Qin, H. Ye, G. Y. Li, and B.-H. Juang, ``Deep learning in physical layer communications,'' \emph{IEEE Wireless Commun.}, vol. 26, no. 2, pp. 93-99, Apr. 2019.

\bibitem{DL2OFDM}
H.~Ye, G.~Y.~Li, and B.-H.~F.~Juang, ``Power of deep learning for channel estimation and
signal detection in OFDM systems,'' \emph{IEEE Wireless Commun. Lett.}, vol.~7, no.~1,
pp. 114-117, Feb. 2018.

\bibitem{CSINet}
C.-K. Wen, W. T. Shih, and S. Jin, ``Deep Learning for Massive MIMO CSI Feedback,'' {\em IEEE Wireless Commun. Lett.},
vol. 7, no. 5, pp. 748-751, Oct. 2018.

\bibitem{DL2018HE}
H.~He, C.-K. Wen, S.~Jin, and G. Y.~Li, ``Deep learning-based channel estimation
for beamspace mmWave massive MIMO systems,'' \emph{IEEE Wireless Commun. Lett.}, vol. 7, no. 5, pp. 852--855, Oct. 2018.

\bibitem{BFNN}
T. Lin and Y. Zhu, ``Beamforming design for large-scale antenna arrays
using deep learning,'' {\em arXiv preprint arXiv:1904.03657,} 2019.

\bibitem{DLbeam}
W. Xia, G. Zheng, Y. Zhu, J. Zhang, J. Wang, and A. P. Petropulu,
``A deep learning framework for optimization of MISO downlink beamforming,'' {\em arXiv preprint, arXiv:1901.00354}, 2019.

\bibitem{C2PO-Net}
A. Balatsoukas-Stimming, O. Castaneda, S. Jacobsson, G. Durisi, and ˜C. Studer, ``Neural-network optimized 1-bit precoding for massive MU-MIMO''
{\em arXiv preprint arXiv:1903.03718,} 2019.

\bibitem{TISTA}
D. Ito, S. Takabe, and T. Wadayama, ``Trainable ISTA for sparse signal
recovery,'' {\em IEEE Trans. Signal. Process.}, vol. 67, no. 12, pp. 3113-3125,
Jun. 2019.

\bibitem{OAMP-Net2}
H. He, C.-K. Wen, S. Jin, and G. Y. Li, ``Model-driven deep learning for MIMO detection,'' {\em IEEE Trans. Signal Process.}, vol. 68,  pp. 1702-1715, Mar. 2020.

\bibitem{MLSP}
M. Y. Takeda, A. Klautau, A. Mezghani, and R. W. Heath, ``MIMO
channel estimation with non-ideal ADCs: Deep learning versus GAMP,''
in {\em Int. Workshop Machine Learning for Signal Process. (MLSP)}, Oct.
2019, pp. 1-6.


\bibitem{stop_gradient}
Y. Bengio, N. Léonard, and A. C. Courville, ``Estimating or propagating gradients through stochastic neurons for conditional computation,''
{\em arXiv preprint arXiv:1308.3432,} 2013.

\end{thebibliography}
\end{document}